\documentclass{article}
\usepackage[utf8]{inputenc}
\usepackage{amsmath}
\usepackage{amsfonts}
\usepackage{amssymb}
\usepackage{amsthm}
\usepackage{xcolor}
 
\usepackage[normalem]{ulem}
\def\bbox#1{{\mathbf{#1}}}

\def\p{\partial}

\theoremstyle{remark}

\newcommand{\dbar}{\bar{\partial}}
\newcommand{\wt}{\widetilde}
\newcommand{\be}{\begin{equation}}
\newcommand{\ee}{\end{equation}}
\newcommand{\bea}{\begin{eqnarray}}
\newcommand{\eea}{\end{eqnarray}}
\newcommand{\beaa}{\begin{eqnarray*}}
\newcommand{\eeaa}{\end{eqnarray*}}

\newcommand{\nn}{\nonumber}

\usepackage{authblk}
\author{L.V. Bogdanov
}
\affil{Landau Institute for Theoretical Physics RAS,
142432 Chernogolovka, Russia}
\author[2]{Lingling Xue}
\affil{Department of Mathematics,
Ningbo University,
Ningbo 315211, P.R. China
}

\title{A class of reductions of the two-component KP hierarchy and the Hirota-Ohta system}
\date{}

\begin{document}
\maketitle
\begin{abstract}
We introduce a class of reductions of the two-component KP hierarchy, which includes 
the Hirota-Ohta system hierarchy.
The description of the reduced hierarchies is based on the Hirota bilinear identity 
and an extra bilinear relation characterising the reduction.
We derive the reduction conditions in terms of the Lax operator and higher linear
operators of the hierarchy, as well as in terms of the basic two-component KP system
of equations.
\end{abstract}
\section{Introduction}
In this work we introduce a class of reductions of the two-component KP hierarchy, which includes 
the Hirota-Ohta system hierarchy \cite{HO}, \cite{Kakei}
as the zero order reduction.
In the scalar case a related class of 
reductions was introduced in \cite{DateVI}, where it was demonstrated that the 
lowest order reductions engender the CKP and BKP hierarchies.
In the two-component case a similar approach was developed
in \cite{BF}.
Our starting point is the $\dbar$-dressing scheme, for which the definition of the class of reduction
is rather transparent \cite{ZM}, and which can be used to construct a big stock of explicit solutions.
However, the algebraic description of the reduced hierarchies is based on the Hirota bilinear identity 
and an 
extra bilinear relation characterising the reduction, and it doesn't necessarily require 
the dressing scheme. 
We derive the reduction conditions in terms of the Lax operator and higher linear
operators of the hierarchy.
The basic system of the two-component KP hierarchy with additional symmetry
constraint for the dynamics defining a special set of times is a closed system of equations with 
three independent variables $x$, $y$, $t$ for
six scalar functions. Each reduction of the class gives a set of three differential relations
containing derivatives with respect to $x$, $y$ for
these functions, and a pair of different reductions produces a closed system of (1+1)-dimensional
equations.
\section{Nonlocal $\dbar$ problem and Hirota bilinear identity}
First we recall a general setting to consider multicomponent KP hierarchy
in the framework of the $\dbar$-dressing method \cite{LVBbook}.
We start from
a pair of adjoint canonically normalised matrix $\dbar$-problems
\begin{eqnarray}
&&
{\partial\over\partial\bar{\lambda}}\chi(\lambda,\bbox{t})
=
\int\!\!\!\int_{\bf C} d\nu\wedge d\bar{\nu}\chi(\nu)g(\nu,\bbox{t})
R(\nu,\lambda)g^{-1}(\lambda,\bbox{t}),\nn\\
&&
{\partial\over\partial\bar{\lambda}}\wt\chi(\lambda,\bbox{t})
=-
\int\!\!\!\int_{\bf C} d\nu\wedge d\bar{\nu}g(\lambda,\bbox{t})R(\lambda,\nu)
g^{-1}(\nu;\bbox{t})\wt\chi(\nu,\bbox{t}).
\label{dbar0}
\end{eqnarray}
We choose
the following parametrization
of the multicomponent loop group ${\Gamma^+}^N$
defining the dynamics of multicomponent KP hierarchy:
\begin{equation}
g(\lambda,\bbox{t})=\exp\left(\sum_{\alpha=1}^N\sum_{n=1}^{\infty}
{P_{\alpha} \lambda^n}t^{(\alpha)}_{n}\right),
\label{g}
\end{equation}
where the
projection matrices $P_{\alpha}$ form a basis of the commutative
subalgebra
of diagonal matrices,
$$
(P_{\alpha})_{\beta\,\gamma}=
\delta_{\alpha\beta}\delta_{\beta\gamma}\quad
(\alpha,\beta,\gamma=1,...,N).
$$
So we have $N$ infinite series of dynamical variables
$t_{(\alpha)\,n}$.

The kernel $R(\lambda,\nu)$ is supposed to be equal to zero in some neighbourhood
of infinity for both variables $\lambda$, $\mu$, for simplicity we suggest that the support of
the kernel belongs to the product of unit disks. Then the functions $\chi(\lambda,\bbox{t})$,
$\wt\chi(\lambda,\bbox{t})$ are analytic outside the unit disc, at
infinity  $\chi(\lambda,\bbox{t})=I + \sum_{n=1}^{\infty} \chi_n(\bbox{t})\lambda^{-n}$,
$\wt\chi(\lambda,\bbox{t})=I + \sum_{n=1}^{\infty} \wt \chi_n(\bbox{t})\lambda^{-n}$
The problems (\ref{dbar0}) imply
Hirota bilinear identity on the unit circle S 
\be
\oint\chi(\nu;\bbox{t})g(\nu,\bbox{t})g^{-1}(\nu,\bbox{t'})
\wt\chi(\nu;\bbox{t'})d\nu=0 
\label{HIROTA00}
\ee
In a more familiar form, for the Baker-Akhieser functions $\psi(\lambda;g)=\chi(\lambda)g(\lambda)$,
$\wt\psi(\lambda;g)=g^{-1}(\lambda)\wt\chi(\lambda)$, we have
\bea
\oint\psi(\nu;\bbox{t})
\wt\psi(\nu;\bbox{t'})d\nu=0. 
\label{HIROTA001}
\eea
We will also use the Cauchy-Baker-Akhieser function (kernel), defined by
nonlocal $\dbar$-problems (\ref{dbar0}) with pole normalisation $(\lambda-\mu)^{-1}$ 
\cite{LVBbook},
\begin{eqnarray}
{\partial\over\partial\bar{\lambda}}\chi(\lambda,\mu;\bbox{t})
=2\pi \text{i} \delta(\lambda-\mu)+
\int\!\!\!\int_{\bf C} d\nu\wedge d\bar{\nu}\;\chi(\nu,\mu;\bbox{t})g(\nu;\bbox{t})
R(\nu,\lambda)g^{-1}(\lambda;\bbox{t}),
&&
\nn\\
{\partial\over\partial\bar{\lambda}}\wt\chi(\lambda,\mu;\bbox{t})
=2\pi \text{i} \delta(\lambda-\mu)-
\int\!\!\!\int_{\bf C} d\nu\wedge d\bar{\nu}\;g(\lambda;\bbox{t}) R(\lambda,\nu)
g^{-1}(\nu;\bbox{t})\wt\chi(\nu,\mu;\bbox{t}).
&&
\label{dbarpole}
\end{eqnarray}
After simple calculations we obtain
\be
\oint \chi(\nu,\lambda;\bbox{t})g(\nu;\bbox{t})g^{-1}(\nu;\bbox{t'})
\wt\chi(\nu,\mu;\bbox{t'})d\nu=0.
\label{HIROTA0}
\ee
It follows from
(\ref{HIROTA0}) taken for $\bbox{t}=\bbox{t'}$
that outside the unit disk with respect to both variables  the function $\chi(\lambda,\mu)$
is equal to $-\wt\chi(\mu,\lambda)$, so in fact this
identity should be written for one function,
\be
\oint \chi(\nu,\lambda;\bbox{t})g(\nu;\bbox{t})g^{-1}(\nu;\bbox{t'})
\chi(\mu,\nu;\bbox{t'})d\nu=0.
\label{HIROTA}
\ee
By similar calculations it is possible to prove that if both problems (\ref{dbarpole}) are solvable,
$\chi(\lambda,\mu)=-\wt\chi(\lambda,\mu)$ for all $\lambda$, $\mu$ where they are commonly
defined. 

Taking  $\lambda\rightarrow\infty$, $\mu\rightarrow\infty$,  we reproduce identity
(\ref{HIROTA00})
for $\chi(\lambda;\bbox{t})=\chi(\lambda,\infty;\bbox{t})$,
$\wt\chi(\lambda;\bbox{t})=-\chi(\infty,\lambda;\bbox{t})$.

In terms of the  Cauchy-Baker-Akhieser (CBA) function 
$$
\Psi(\lambda,\mu;\bbox{t})=g^{-1}(\mu,\bbox{t})\chi(\lambda,\mu;\bbox{t})g(\lambda,\bbox{t})
$$
the Hirota bilinear identity reads
\be
\oint \Psi(\nu,\lambda;\bbox{t})
\Psi(\mu,\nu;\bbox{t'})d\nu=0.
\label{HIROTA_CBA}
\ee
\section{A class of reductions of the two-component KP hierarchy}
For the two-component KP hierarchy
\begin{equation}
g(\lambda,\bbox{t})=\exp\left(\sum_{n=1}^{\infty}
({P_{1} \lambda^n}t^{(1)}_{n} + {P_{2} \lambda^n}t^{(2)}_{n})
\right),
\label{2g}
\end{equation}
We consider a class of reductions 
\bea
 R^\text{T}(-\lambda,-\mu)A\mu^k=A\lambda^k J R(\mu,\lambda) J^{-1},
\label{reductions}
\eea
where
$
J=
\begin{pmatrix}
0&1\\
-1&0
\end{pmatrix},
$
the matrix $A$ is equal to $I$ or $\sigma_3$, so it commutes or anticommutes
with $J$.
This class of reduction requires an involution
\bea
g(-\lambda,\bbox{t})=J g(\lambda,\bbox{t})^{-1} J^{-1},
\label{involution}
\eea
so the reduction condition is compatible with the dynamics only if
$t^{(1)}_{2n+1}= t^{(2)}_{2n+1}$ (for odd  times) 
and $t^{(1)}_{2n}=- t^{(2)}_{2n}$ (for even times).
We introduce a new set of times
$t_n$, for even order $t_{2k}=t^{(1)}_{2k}=- t^{(2)}_{2k}$
and for odd order $t_{2k-1}=t^{(1)}_{2k-1}= t^{(2)}_{2k-1}$.
The factor definig the dynamics of the kernel looks like
\begin{equation}
g(\lambda,\bbox{t})=\exp\left(\sum_{n=1}^{\infty}
(I \lambda^{2n-1}t_{2n-1} + \sigma_3 \lambda^{2n} t_{2n})
\right).
\label{2gR}
\end{equation}
For the first three times we will use the notations $x=t_1$,
$y=t_2$, $t=t_3$.

In terms of the Baker-Akhiezer function the reduction (\ref{reductions}) is 
characterised by an extra
bilinear relation
\be
\oint\psi(\nu;\bbox{t})JA\nu^{k}
\psi^\text{T}(-\nu;\bbox{t'})d\nu=0,
\label{HIROTAred1}
\ee
for the CBA function we have 
\bea
\oint\Psi(\nu,\lambda;\bbox{t})JA\nu^{k}
\Psi^\text{T}(-\nu,-\mu;\bbox{t'})d\nu=0.
\label{HIROTAred2}
\eea
Reduction condition corresponding to the Hirota-Ohta hierarchy is given
by (\ref{reductions}) with $A=I$, $n=0$,
\beaa
 R^\text{T}(-\lambda,-\mu)=J R(\mu,\lambda) J^{-1},
\eeaa
for the Baker-Akhiezer functions we have a condition 
\be
\wt\psi(\lambda;\bbox{t})=-J\psi^\text{T}(-\lambda;\bbox{t})J,
\label{BAred}
\ee
and identity (\ref{HIROTAred1}) reads 
\be
\oint\psi(\nu;\bbox{t})J
\psi^\text{T}(-\nu;\bbox{t'})d\nu=0. 
\label{HIROTAred2}
\ee
In terms of the the CBA function the reduction condition is
\beaa
 \psi^\text{T}(-\lambda,-\mu)=J \psi(\mu,\lambda) J^{-1}.
\eeaa
\section{Two-component KP hierarchy}
Having in mind the class of reductions (\ref{reductions}), we will first derive linear problems
and equations for the two-component KP hierarchy case with the 
involution  (\ref{involution}), and then
we will consider reduction conditions in terms of linear operators and equations.

Let us start with Hirota bilinear identity (\ref{HIROTA0}), (\ref{HIROTA}) with 
the dependence on times defined
by  (\ref{2gR}).

\subsection*{Linear operators}
The action of operators $\p_{t_n}=\frac{\p~}{\p t_n}$ on $\psi$, $\psi^\ast=\wt \psi^\text{T}$
corresponds to the following operators (the Manakov operators)
acting on $\chi$, $\chi^*$
\beaa
&&
D_{t_{2n-1}}\chi=\p_{t_{2n-1}}\chi + \lambda^{2n-1}\chi, \quad 
D^\ast_{t_{2n-1}}\chi=\p_{t_{2n-1}}\chi^\ast - \lambda^{2n-1}\chi^\ast;
\\
&&
D_{t_{2n}}\chi=\p_{t_{2n}}\chi + \lambda^{2n}\chi \sigma_3, \quad 
D^\ast_{t_{2n}}\chi=\p_{t_{2n}}\chi^\ast - \lambda^{2n}\chi^\ast\sigma_3;
\eeaa
Hirota bilinear identity implies that some differential operator 
$\sum_{n,m} u^{(n)}_{m} \p_n^m$ acting on $\psi$ gives zero iff
for respective Manakov operator the result of action on $\chi$ has zero projection 
to nonnegative powers of $\lambda$,
\bea
(\sum_{n,m} u^{(n)}_{m} D_n^m\chi)_+=0 
\Leftrightarrow
\sum_{n,m} u^{(n)}_{m} D_n^m\chi=0
\Leftrightarrow
\sum_{n,m} u^{n}_{m} \p_n^m\psi=0
\label{crit}
\eea
Using this observation, it is possible to construct linear operators of the hierarchy. 

Let us start with the Lax operator. For the first three times  $x=t_1$,
$y=t_2$, $t=t_3$ the Manakov operators look like
\beaa
&&
D_x\chi=\p_x\chi + \lambda\chi,\quad
D_y\chi=\p_x\chi + \lambda^2\chi \sigma_3,\quad
D_t\chi=\p_x\chi + \lambda^3\chi,\\
&&
D^\ast_x\chi^\ast=\p_x\chi^\ast - \lambda\chi^\ast,\quad
D^\ast_y\chi^\ast=\p_x\chi^\ast - \lambda^2\chi^\ast \sigma_3,\quad
D^\ast_t\chi^\ast=\p_x\chi^\ast - \lambda^3\chi^\ast,
\eeaa
Using (\ref{crit}), we obtain
\beaa
&&
(D_y-\sigma_3 D_x^2)\chi=\left(
[\chi_1,\sigma_3]D_x +
(
[\chi_2,\sigma_3]-2\sigma_3\chi_{1x}- [\chi_1,\sigma_3]\chi_1
)
\right)\chi,
\label{L0}
\\
&&
(D^*_y+\sigma_3 {D^*_x}^2)\chi^*=
\left(
[\chi^*_1,\sigma_3]D^*_x -
(
[\chi^*_2,\sigma_3]+2\sigma_3\chi^*_{1\,x} - [\chi^*_1,\sigma_3]\chi^*_1
)
\right)\chi^*.
\eeaa
Thus for the Baker-Akhiezer functions 
\bea
&&
\p_y \psi=
\left(
\sigma_3\p_x^2 + 
\begin{pmatrix}
0&f\\
g&0
\end{pmatrix}\p_x
+U
\right)
\psi,
\label{L}\\
&&
\p_y \psi^*=
\left(
-\sigma_3\p_x^2 +
\begin{pmatrix}
0&f^*\\
g^*&0
\end{pmatrix}\p_x
-U^*
\right)
\psi^*,
\nn
\eea
where
\beaa
\begin{pmatrix}
0&f\\
g&0
\end{pmatrix}
=[\chi_1,\sigma_3],\quad
\begin{pmatrix}
0&f^*\\
g^*&0
\end{pmatrix}=-[\chi^*_1,\sigma_3],
\\
U=[\chi_2,\sigma_3]-2\sigma_3\chi_{1x}- [\chi_1,\sigma_3]\chi_1,
\\
U^*=[\chi^*_2,\sigma_3]+2\sigma_3\chi^*_{1\,x} -  [\chi^*_1,\sigma_3]\chi^*_1.
\eeaa
From Hirota identity taken for equal times we get
\bea
\oint\chi(\nu;\bbox{t})
{\chi^*}^\text{T}(\nu;\bbox{t})d\nu=0,
\eea
then $\chi^*_1=-\chi_1^{\text{T}}$ and $g=f^*$, $f=g^*$.
Differentiating identity (\ref{HIROTA00}) 
with respect to $x$
and taking it for equal times, 
we get 
\bea
\oint\chi^*(\nu;\bbox{t})
(\p_x+\nu)\chi^\text{T}(\nu;\bbox{t})d\nu=0,
\eea
that implies 
\beaa
\chi_2^*(\bbox{t})=-\chi^\text{T}_2 - \chi^\text{T}_{1\,x} + \chi^\text{T}_1\chi^\text{T}_1.
\eeaa
Using this relation it is easy to demonstrate that L-operators (\ref{L}) are (anti)adjoint 
(defining $(V\p_x)^*=-\p_x V^\text{T}$),
\bea
\p_y \psi=\bbox{B}_2 \psi, \quad \p_y \psi^*=-\bbox{B}_2^* \psi^*,
\label{B2}
\eea
where the operator $\bbox{B}_2$ is defined by (\ref{L}).

Linear operators corresponding to the time $t$ read
\bea
\p_t \psi=\bbox{B}_3 \psi, \quad \p_t \psi^*=-\bbox{B}_3^* \psi^*,
\label{B3}
\eea
where
\beaa
&&
\bbox{B}_3=\p_x^3 + 3 W\p_x + W_1,
\\
&&
W= - \chi_{1x},
\quad W_1=3\chi_{1x}\chi_1-3\chi_{2x}-3\chi_{1xx}.
\eeaa
Higher linear operators can be written as
\bea
\p_t \psi=\bbox{B}_n \psi, \quad \p_t \psi^*=-\bbox{B}_n^* \psi^*,
\label{Bn}
\eea
where
\beaa
&&
\bbox{B}_{2m}=\sigma_3\p_x^{2m}  + 
\begin{pmatrix}
0&f\\
g&0
\end{pmatrix}
\p^{2m-1}_x + 
\sum_{k=0}^{2n-2} U^{(2m)}_k \p_x^k,
\\
&&
\bbox{B}_{2m+1}=\p_x^{2m+1}  + (2m+1) W\p^{2m-1}_x + 
\sum_{k=0}^{2n-2} W^{(2m+1)}_k \p_x^k
\eeaa
The coefficients of these operators can be expressed through the coefficients of
expansion of the function $\chi(\lambda;\bbox{t})$. The Lax operator (\ref{L})
written in terms of this function gives  the recursion formulae, expressing the coefficients 
of expansion $\chi_n(\bbox{t})$ through the coefficients of the Lax operator
$f$, $g$ and $U$ (six scalar functions).
Indeed, 
\beaa
\chi_y-\sigma_3\chi_{xx}=
-\lambda^2[\chi,\sigma_3]  + 2\lambda \sigma_3\chi_x
+
\begin{pmatrix}
0&f\\
g&0
\end{pmatrix}
(\chi_x + \lambda \chi) +U\chi,
\eeaa
and in terms of coefficients of expansion  we have
\beaa
\chi_{ky}-\sigma_3\chi_{kxx}=
-[\chi_{k+2},\sigma_3]  + 2\sigma_3\chi_{k+1\,x}
+
\begin{pmatrix}
0&f\\
g&0
\end{pmatrix}
(\chi_{kx} + \chi_{k+1}) +U\chi_k.
\eeaa
To have a correct structure of recursion, it is necessary to split this equation
into diagonal and antidiagonal part,
then we obtain
\bea
2\chi^a_{k+2}= - 2\chi^a_{k+1\,x} +
\begin{pmatrix}
0&-f\\
g&0
\end{pmatrix}
(\chi^d_{k+1} +\chi^d_{k x}) - \sigma_3 (U\chi_k)^a + \sigma_3\chi^a_{k\,y} -\chi^a_{k\,xx},
&&
\label{a}\\
2\chi^d_{k+1\,x}=
\begin{pmatrix}
0&-f\\
g&0
\end{pmatrix}
(\chi^a_{k+1} +\chi^a_{k x}) - \sigma_3 (U\chi_k)^d  + \sigma_3\chi^d_{k\,y} -\chi^d_{k\,xx}.
\qquad
&&
\label{d}
\eea
Let us write down several terms of the recursion explicitly:
\\
anti-diagonal part, $k=-1$,
\beaa
2\chi^a_{1}=\begin{pmatrix}
0&-f\\
g&0
\end{pmatrix},
\eeaa
diagonal part, $k=0$,
\beaa
2\chi^d_{1\,x}=-\frac{1}{2}fg I - \sigma_3 U^d,
\eeaa
anti-diagonal part, $k=0$,
\beaa
2\chi^a_{2}=-2\chi^a_{1\,x} + \begin{pmatrix}
0&f\\
-g&0
\end{pmatrix}
\chi^d_{1} 
- \sigma_3 U^a,
\eeaa
diagonal part, $k=1$,
\beaa
2\chi^d_{2\,x}=
\begin{pmatrix}
0&f\\
-g&0
\end{pmatrix}
(\chi^a_{2} +\chi^a_{1 x}) - \sigma_3 (U\chi_1)^d + \sigma_3\chi^d_{1\,y} -\chi^d_{1\,xx}.
\eeaa
\subsection*{Equations}
Compatibility condition for linear equations (\ref{B2}), (\ref{B3})
is given by the Zakharov-Shabat equation
\beaa
\frac{\p\bbox{B}_3}{\p y}-\frac{\p \bbox{B}_2}{\p t}
= \left[ \bbox{B}_2, \bbox{B}_3 \right] , 
\eeaa
which engenders a closed system of equations for  matrix coefficients
of the operators $\bbox{B}_2$, $\bbox{B}_3$:
\bea
&&
U_t - W_{1y} - U_{xxx} + \sigma_3 W_{1xx} + 
\begin{pmatrix}
0&f\\
g&0
\end{pmatrix}
W_{1x} +[U,W_1]  -3WU_x=0,
\nn\\
&&
\begin{pmatrix}
0&f_t\\
g_t&0
\end{pmatrix}
+ 3\sigma_3 W_{xx} 
- 3W
\begin{pmatrix}
0&f_x\\
g_x&0
\end{pmatrix}
+3
\begin{pmatrix}
0&f\\
g&0
\end{pmatrix}
W_x
-3[W,U] 
\nn\\
&&\qquad
- [W_1, 
\begin{pmatrix}
0&f\\
g&0
\end{pmatrix}
]+2\sigma_3W_{1x} -3 W_y -3 U_{xx}
=0,
\nn\\
&&
3U_x + 3
\begin{pmatrix}
0&f_{xx}\\
g_{xx}&0
\end{pmatrix} 
-6\sigma_3 W_x +[W_1,\sigma_3]
+3[W,
\begin{pmatrix}
0&f\\
g&0
\end{pmatrix}
]=0,
\nn\\
&&
\begin{pmatrix}
0&f_x\\
g_x&0
\end{pmatrix}
=
[W,\sigma_3].
\label{System}
\eea
This system represents a two-component KP system for the times $x$, $y$, $t$.
Having in mind the recursion relations (\ref{a}), (\ref{d}) and expressions for
$W$, $W_1$ (\ref{B3}), we come to the conclusion that all the matrix functions in this
system can be expressed through $f$, $g$, $U$, and the system should 
give a closed system of equations for six scalar functions.

\section{Reductions}
\subsection*{Hirota-Ohta system hierarchy}
The Hirota-Ohta system hierarchy is a reduction of the two-component KP hierarchy
defined by the condition (\ref{BAred}), which is equivalent to
\beaa
\psi^*(\lambda;\bbox{t})=-J\psi(-\lambda;\bbox{t})J.
\eeaa
Then for linear operators (\ref{Bn}) we have
$
\bbox{B}_n^*=J\bbox{B}_nJ
$
(compare \cite{Kakei}), and the reduced operator $\bbox{B}_2$ (\ref{L}) is of the form
\bea
&&
\bbox{B}_2=
\sigma_3\p_x^2 + 
2
\begin{pmatrix}
u&v\\
-\wt v &-u
\end{pmatrix}
\psi,
\label{L_HO}
\eea
$f=g=0$, $\chi_1=-u I$. 
For reduced operator $\bbox{B}_3$ $W=uI$. Reduced system (\ref{System}) reads
\beaa
&&
U_t - W_{1y} - U_{xxx} + \sigma_3 W_{1xx}  
+[U,W_1]  -3uU_x=0,
\\
&&
3\sigma_3 u_{xx} 
+2\sigma_3W_{1x} -3I u_y -3 U_{xx}
=0,
\\
&&
3U_x 
-6\sigma_3 u_x +[W_1,\sigma_3]
=0,
\eeaa
where the second equation gives the expression for $W_{1x}$ in terms of $U$,
the third equation is implied by the second. From the first equation we get
an equation for one matrix $U$ of the form 
$U=u\sigma_3 + U^{\text{a}}$ (antidiagonal part), which, written in components, represents 
the Hirota-Ohta (coupled KP) system  \cite{HO}, \cite{Kakei}
\bea
&& \left(4u_t -u u_x -12u_{xxx}\right)_x 
-3u_{yy}+12(v\tilde{v})_{xx}=0, \nonumber\\
&& 2v_t + 6u v_x + v_{xxx} + 3v_{xy} + 6v \p_x^{-1} u_y = 0,
\label{cKP}\\
&& 2\tilde{v}_t + 6u\tilde{v}_x +\tilde{v}_{xxx}
- 3\tilde{v}_{xy} - 6\tilde{v}\p_x^{-1} u_y  = 0. \nonumber
\eea
\subsection*{Other reductions}
Let us consider another zero order  reduction (\ref{reductions}), (\ref{HIROTAred1}) 
with $A=\sigma_3$, 
$k=0$. Effectively that leads to the change of the matrix $J$ to the matrix 
$$
J'=\sigma_3 J
=\begin{pmatrix}
0&1\\
1&0
\end{pmatrix}.
$$
Then 
\beaa
\psi^*(\lambda;\bbox{t})=J'\psi(-\lambda;\bbox{t})J',
\eeaa
for linear operators (\ref{Bn}) we have
$
\bbox{B}_n^*=-J'\bbox{B}_nJ'.
$
The reduced Lax operator $\bbox{B}_2$ (\ref{L}) is of the form
\beaa
&&
\p_y \psi=
\left(
\sigma_3\p_x^2 + 
\begin{pmatrix}
0&f\\
g&0
\end{pmatrix}\p_x
+uI
+ \frac{1}{2}\begin{pmatrix}
0&f_x\\
g_x&0
\end{pmatrix}
\right)
\psi.
\eeaa
Thus the Lax operator depends on three functions $f$, $g$, $u$ instead of six functions
in the general case of the two-component KP hierarchy,
and the matrix system (\ref{System}) should give a closed system of equations for
these three functions. 
\subsubsection*{First order reductions}
Let us consider the reduction (\ref{HIROTAred1}) with $A=I$, 
$k=1$,
\beaa
\oint\psi(\nu;\bbox{t})J\lambda
\psi^\text{T}(-\nu;\bbox{t'})d\nu=0. 
\eeaa
Taking this identity for equal times, we get
\beaa
\chi_2 J+J\chi_2^{\text{T}} - \chi_1 J \chi_1^{\text{T}}=0
\eeaa
Recalling the recursion relations (\ref{a}), (\ref{d}),  we obtain 
three scalar differential relations for six functions $f$, $g$, $U$.
These relations represent a reduction for the system (\ref{System}).
\subsubsection*{Higher reductions}
Higher reductions may be considered in a similar way.
In general, a reduction of arbitrary order (or adjoint reduction) represents
a set of three scalar differential relations for six functions $f$, $g$, $U$.
A pair of reductions of different orders engenders a closed (1+1)-dimesional
system for six functions, connected with some stationary reductions of the hierarchy.

Another way to characterise the reduction is the existence of intertwining
differential operator $\bbox{A}_k$ of the order $k$, which defines a map from the
wave functions of adjoint operators to the wave functions of basic linear operators.
Similar idea was used in \cite{BF} 
 to construct the differential reductions for the
case of the 
two-dimensional Dirac operator.
It is convenient to introduce a modified conjugation operation,
for matrix differential operator $\bbox{B}$ we define $\bbox{B}^\dag=J\bbox{B}^*J^{-1}$.
This operation possesses standard properties
$(\bbox{B}^\dag)^\dag=B$, $(\bbox{A}\bbox{B})^\dag=\bbox{B}^\dag\bbox{A}^\dag$.
We denote $\psi^\dag(\lambda;\bbox{t})=J\psi^*(\lambda;\bbox{t})J^{-1}$.
Then the reduction is characterised by the existence of differential operator $\bbox{A}_k$ 
of the order $k$, such that for any wave function~$\phi$
\beaa
(\p_y +\bbox{B}^\dag_2)\phi=0\Rightarrow (\p_y - \bbox{B}_2)\bbox{A}_k\phi=0.
\eeaa
Algebraically, this condition is equivalent to the operator equation
\bea
(\p_y -\bbox{B}_2)\bbox{A}_k=  \bbox{A}_k(\p_y + \bbox{B}^\dag_2) ,
\label{operator}
\eea
see \cite{DateVI} for the scalar case. Using this equation, it is possible to express 
the coefficients of operator $\bbox{A}_k$ through the coefficients of the Lax operator
and get a reduction condition in terms of the coefficients of the Lax operator
(or the solution of the system (\ref{System})).
This type of condition can be also used 
to define the reductions
in the context of the Lax-Sato equations (the scalar case is considered
in \cite{DateVI}).
\section{Appendix. Reductions in terms of the Lax-Sato equations}
Here we briefly describe the Lax-Sato picture of the two-component KP hierarchy 
with the times (\ref{2gR})  \cite{Kakei} and of the class of reductions corresponding to the
bilinear relation (\ref{HIROTAred1}). In the scalar case reductions of this type are described
in \cite{DateVI}.

The Lax-Sato equations define the dynamics of pseudodifferential operators 
\beaa
&&
L=\partial + U_1\partial^{-1}  + U_2\partial^{-2} + \dots,
\\
&&
M=\sigma_3 + V_1\partial^{-1} + V_2\partial^{-2} + \dots,
\eeaa
where $U_n$, $V_n$ are $2\times 2$ matrices, $\p=\p_x$,
$
\sigma_3=
\begin{pmatrix}
1&0\\
0&-1
\end{pmatrix}
$,
with the characteristic properties
\beaa
[L,M]=0, 
\qquad
M^2=1
\eeaa
For odd times:
\bea
\frac{\p L~~~}{\p t _{2n+1}}=[(L^{2n+1})_+, L],
\qquad
\frac{\p M~~~}{\p t _{2n+1}}=[(L^{2n+1})_+, M],
\label{LModd}
\eea
for even times:
\bea
\frac{\p L~~}{\p t _{2n}}=[(L^{2n}M)_+, L],
\qquad
\frac{\p M~}{\p t _{2n}}=[(L^{2n}M)_+, M]
\label{LMeven}
\eea
The Gelfand-Dickey reductions for this hierarchy are defined by the following conditions:
for odd flows
\beaa
(L^{2n+1})_-=0,\quad (L^{2n+1})_+= \bbox{D}^{(2n+1)},
\eeaa
where $D^{(2n+1)}$ is a differential operator of the order $2n+1$
with matrix coefficients .
For even flows:
\beaa
(L^{2n}M)_-=0,\quad (L^{2n}M)_+= \bbox{D}^{(2n)}
\eeaa
Introducing formal pseudodifferential 
dressing operator (connected to operator $(1+\hat K_+)$ used by Kakei)
\beaa
P=I + W_1\partial^{-1}  + W_2\partial^{-2} + \dots,
\eeaa
it is possible to express operators $L$ and $M$ as
\beaa
L=P\p P^{-1},
\quad
M=P \sigma_3 P^{-1}
\eeaa
The operators $L$ and $M$  defined this way evidently possess necessary
characteristic properties.
Dynamics of the dressing operator is defined by the Sato equations \cite{Sato}, 
\bea
&&
\frac{\p P~~~}{\p t _{2n+1}}=- (P\p^{2n+1} P^{-1})_- P,
\nn\\
&&
\frac{\p P~~}{\p t _{2n}}=-(P\p^{2n}\sigma_3 P^{-1})_-P,
\label{Sato}
\eea
that implies (\ref{LModd}), (\ref{LMeven}).
To find a dressing operator starting from $L$ (or $M$), one should solve
a factorization problem
$LP=P\p$,
$MP=P\sigma_3$.
Reduction to the Hirota-Ohta system hierarchy is described by the conditions
\beaa
&&
L^*=JLJ,\\
&&
M^*=JMJ,\\
&&
P^*=-JP^{-1}J.
\eeaa
A class of reductions corresponding to bilinear relation (\ref{HIROTAred1}) 
is defined by the conditions:
for $A=I$
\beaa
(P\p^n J P ^*)_-=0,
\eeaa
for  $A=\sigma_3$
\beaa
(P\sigma_3 J\p^n P ^*)_-=0.
\eeaa
Introducing the differential operators $\bbox{A}_k=P\p^{k} P^\dag$ (for $A=I$) or
$\bbox{A}_k=P\sigma_3\p^{k} P^\dag$ (for  $A=\sigma_3$), where we use the notation
$P^\dag=J P^*J^{-1}$,
we obtain the relations
\beaa
&&
L \bbox{A}_k=\bbox{A}_k L^{\dag},\\
&&
M\bbox{A}_k=\bbox{A}_k M^{\dag},
\eeaa
and also relations of the form (\ref{operator}).

\subsection*{Acknowledgements}The reported study was funded by RFBR and NSFC, 
project number 21-51-53017.
\subsection*{Conflicts of Interest} 
The authors declare that they have no conflicts of interests.

\end{document}